\documentclass[pre,twocolumn, superscriptaddress, a4paper, showpacs]{revtex4}
\usepackage{amssymb,amsfonts,amsmath}
\usepackage{graphicx}

\begin{document}
\title{Continuous time multidimensional Markovian description of L\'evy walks}
\author{Ihor Lubashevsky}
    \affiliation{A.M. Prokhorov General Physics Institute, Russian
    Academy of Sciences, Vavilov Str. 38, 119991 Moscow, Russia}
    \affiliation{Institut f\"ur Physikalische Chemie,
    Westf\"alische Wilhelms Universit\"at M\"unster,  Corrensstr. 30,
    48149 M\"unster, Germany}
\author{Rudolf Friedrich}
    \affiliation{Institut f\"ur Theoretische Physik,
    Westf\"alische Wilhelms Universit\"at M\"unster, Wilhelm-Klemm. 9, 48149 M\"unster,
    Germany}
    \affiliation{Center of Nonlinear Science CeNoS,
    Westf\"alische Wilhelms Universit\"at M\"unster, 48149 M\"unster, Germany}
\author{Andreas Heuer}
    \affiliation{Institut f\"ur Physikalische Chemie, Westf\"alische Wilhelms
    Universit\"at M\"unster,  Corrensstr. 30,
    48149 M\"unster, Germany}
    \affiliation{Center of Nonlinear Science CeNoS,
    Westf\"alische Wilhelms Universit\"at M\"unster, 48149 M\"unster, Germany}
\date{\today}
\begin{abstract}

The paper presents a multidimensional model for nonlinear Markovian random walks that generalizes one we developed previously (Phys.~Rev.~E~\textbf{79},~011110,~2009) in order to describe the L\'evy type stochastic processes in terms of continuous trajectories of walker motion. This approach may open a way to treat L\'evy flights or L\'evy random walks in inhomogeneous media or systems with boundaries in the future.
The proposed model assumes the velocity of a wandering particle to be affected by a linear friction and a nonlinear Langevin force whose intensity is proportional to the magnitude of the velocity for its large values. Based on the singular perturbation technique the corresponding Fokker-Planck equation is analyzed and the relationship between the system parameters and the L\'evy exponent is found. Following actually the previous paper we demonstrate also that anomalously long displacements of the wandering particle are caused by extremely large fluctuations in the particle velocity whose duration is determined by the system parameters rather than the duration of the observation interval. In this way we overcome the problem of ascribing to L\'evy random walks non-Markov properties.

\end{abstract}


\maketitle

\section{Introduction}

There is a wide class of physical systems, where transport phenomena exhibit anomalous behavior of the L\'evy type called also superdiffusion (for a review, see, e.g., Ref.~\cite{{nmlf3}}). For such processes a variable $\mathbf{x}$ describing, in particular, spatial displacement of a wandering particle in the $N$-dimensional space $\mathbb{R}^N$ during time interval $t$ is characterized by some scale $\ell(t)$ increasing with time $t$ as $\ell(t)\propto t^{1/\alpha}$, where the parameter $\alpha$ meets the inequality $\alpha < 2$. In addition, the distribution function $P(\mathbf{x},t)$ of the particle displacement has the asymptotic behavior $\ln\left[P(\mathbf{x},t)\right] \sim -\ell(t)^\alpha/|\mathbf{x}|^{N+\alpha}$ for $|\mathbf{x}|\gg\ell(t)$.

In particular, it is the case for the motion of tracer particles in turbulent flows \cite{Swinney}, the diffusion of particles in random media \cite{Bouchaud}, human travel behavior and spreading of epidemics \cite{Brockmann} or economic time series in finance \cite{Stanley}. Recently, there has been a great deal of research about superdiffusion. It includes, in particular, a rather general analysis of the Langevin equation with L\'evy noise (see, e.g., Ref.~\cite{Weron}) and the form of the corresponding Fokker-Planck equations \cite{Schertzer1,Schertzer2,CiteNew1,CN100}, description of anomalous diffusion with power law distributions of spatial and temporal steps \cite{Fogedby1,Sokolov}, L\'evy flights in heterogeneous media \cite{Fogedby2,Honkonen,BrockmannGeisel} and in external fields \cite{BrockmannSokolov,Fogedby3}, constructing the Fokker-Planck equation for L\'evy type processes in nonhomogeneous media~\cite{CiteNew2,CiteNew3,CiteNew4}, first passage time analysis and escaping problem for L\'evy flights \cite{fptp1,fptp2,fptp3,fptp4,fptp5,fptp6,CiteNew5,CiteNew6}, as well as processing experimental data for detecting the L\'evy type behavior~\cite{SiegertLevy}. Besides, it should be noted that the attempt to consider L\'evy flights in bounded systems (see, e.g., Ref.~\cite{nmlf1,nmlf2} and review~\cite{nmlf3} as well) has introduced the notion of L\'evy walks being a non-Markovian process because of the necessity to bound the walker velocity. The problems arise because for a  L\'evy walk the second moment $\left<(\mathbf{x}_t-\mathbf{x}_i)^2\right>$ ($\mathbf{x}_i$ and $\mathbf{x}_t$ are the initial and final point, respectively) diverges.

Previously, we developed an one-dimensional model that generates continuous Markovian trajectories, following the L\'evy statistics, by using simple Gaussian but multiplicative noise for the time evolution of the velocity \cite{We}. The spatial dynamics naturally follows from this. It should be pointed out that a first step in this direction can be found in Refs.~\cite{Sak01,Fra04}. For a fixed time scale $\delta t$ we can recover the standard behavior of L\'evy type processes. However, we have full locality in the sense that a trajectory can be determined with any desired resolution.
In other words, the developed model proposes a microscopic implementation of the L\'evy type processes characterized by an arbitrary small time scale $\tau$ that can be chosen beforehand. When running time exceeds essentially this microscopic time scale, $t\gg\tau$, the corresponding random walks, as should be, are described by the L\'evy distribution. 

The purpose of the present paper is, first, to generalize the developed one-dimensional model to multidimensional case and, second, to obtain the rigorous results using an original singular perturbation technique. The previous paper \cite{We} was actually devoted to the formulation of the problem at hand and the qualitative explanation of its main properties using numerical simulation. In the present paper the one-dimensional model is also analyzed in detail as a specific limit case and the corresponding results just declared previously are obtained in a rigorous way.  The approach to be developed may open in the future a way to treat L\'evy flights or L\'evy random walks in inhomogeneous media or systems with boundaries including boundaries of complex geometry.

\section{Stochastic system and the governing equations}

We consider continuous random walks in the Euclidean $N$-dimensional space $\mathbb{R}^{N}$ governed by the following stochastic differential equations of the H\"{a}nggi-Klimontovich type (a post-point process in the given case) \cite{H1,H2,Kl}
\begin{align}
\label{2Klim1}
     \frac{dx_{i}}{dt}  &  =v_{i}\,,\\
\label{2Klim2}
     \frac{dv_{i}}{dt}  &  =-\frac{(N+\alpha)}{\tau}v_{i}+\sqrt{\frac{2}{\tau}}\,g(\mathbf{v})\ast\xi_{i}(t)\,,
\end{align}
for $i=1,2,\ldots,N$. Here the symbol $*$ denotes the belonging of the given product to the H\"{a}nggi-Klimontovich type,   $\mathbf{x}=\{x_{1,}x_{2},\ldots,x_{N}\}$ is the point of $\mathbb{R}^{N}$, the vector $\mathbf{v}=\{v_{1,}v_{2},\ldots,v_{N}\}$ is the current particle velocity, and the collection of mutually independent random variables $\{\xi_{i}(t)\}_{i=1}^{i=N}$  such that
\begin{equation}\label{3}
      \left\langle \xi_{i}(t)\right\rangle =0\quad\text{and}\quad\left\langle \xi_{i}(t)\xi_{i^{\prime}}(t^{\prime})\right\rangle
     =\delta_{ii^{\prime}}\delta(t-t^{\prime})
\end{equation}
represents the Langevin forces with the amplitude
\begin{equation}\label{4}
     g(\mathbf{v})=\sqrt{v_a^{2}+\mathbf{v}^{2}}\,.
\end{equation}
The dimensionless coefficient $\alpha>0$, the time scale $\tau$, and the characteristic velocity $v_a$ are the system parameters. According to the
results to be obtained, the Lev\'{y} type behavior can be found for
\begin{equation}\label{alpha}
     1<\alpha<2\,.
\end{equation}
Exactly this region of $\alpha$ will be considered below. The scalar form of the coefficient $\alpha$ and the Langevin force intensity~\eqref{4} actually implement the adopted assumption about the isotropy of the system at hand. For $N=1$ the present model coincides with one developed in Ref.~\cite{We} within the replacement $\tau\to 2\tau$.

For the given system the distribution function $\mathcal{P}\left(\mathbf{x}-\mathbf{x}_{0},\mathbf{v},\mathbf{v}_{0},t\right)$ obeys the following Fokker-Planck equation written in the kinetic form
\begin{multline}\label{FPKlim}
  \frac{\partial\mathcal{P}}{\partial t}=\sum_{i=1}^{N}\left\{-\frac{\partial}{\partial x_{i}}\left[  v_{i}\mathcal{P}\right]\right.
\\
   \left.
  {}+\frac1{\tau}\frac{\partial}{\partial v_{i}}\left[  g^{2}(\mathbf{v})\frac{\partial\mathcal{P}}{\partial v_{i}}
  +(N+\alpha)v_{i}\mathcal{P}\right]  \right\}
\end{multline}
subject to the initial condition at $t=0$
\begin{equation}\label{FP2}
  \mathcal{P}\left(  \mathbf{x}-\mathbf{x}_{0},\mathbf{v},\mathbf{v},0\right)
  =\delta(\mathbf{r}-\mathbf{r}_{0})\delta(\mathbf{v}-\mathbf{v}_{0})\,,
\end{equation}
where, an addition, the system translation invariance with respect to the variable $\mathbf{x}$ is taken into account explicitely.

\subsection*{Essence of the proposed description}

Let us discuss the core idea of the mathematical description of the L\'evy type random walks within the continuous time Markovian process. In the general form the corresponding Fokker-Planck equation~\eqref{FPKlim} can be rewritten as 
\begin{equation}\label{Fr0}
   \frac{\partial \mathcal{P}}{\partial t} = \left[-\mathbf{v}\cdot\nabla_{\mathbf{x}} + \widehat{L}_{\mathbf{v}}\right]\mathcal{P}
\end{equation}
where the operator $ \widehat{L}_{\mathbf{v}}$ acting on the variable $\mathbf{v}$ only is given by the second term on the right hand side of expression~\eqref{FPKlim}. Since our analysis is confined to the long time dynamics of the variables $\mathbf{x}$ the solution $\mathcal{P}$ of this equation is sought in the form of the expansion
\begin{equation}\label{Fr1}
  \mathcal{P}=\sum_\Lambda \varPsi_\Lambda\left(\mathbf{v},\nabla_\mathbf{x}\right) f_\Lambda(\mathbf{x},t)
\end{equation}
over the eigenfunctions $\varPsi_\Lambda\left(\mathbf{v},\nabla_\mathbf{x}\right)$ matching the suitably defined eigenvectors $\Lambda\left(\nabla_\mathbf{x}\right)$ of the linear eigenvalue problem
\begin{equation}\label{Fr2}
  \Lambda\left(\nabla_\mathbf{x}\right) \left(\mathbf{v},\nabla_\mathbf{x}\right)  =
  \left[-\mathbf{v}\cdot\nabla_{\mathbf{x}} + \widehat{L}_{\mathbf{v}}\right]
  \varPsi_\Lambda \left(\mathbf{v},\nabla_\mathbf{x}\right)\,,
\end{equation}
where the operator $\nabla_\mathbf{x}$ is treated as some formal parameter. We expect that the long time dynamics will be described only by a few terms of such expansion. These eigenfunctions are assumed to be complete and form a basis, which is justified as will be seen below. Furthermore, according to the general properties of such random walks the zeroth approximation of the eigenvalue problem~\eqref{Fr2} determined by the replacement $\nabla_\mathbf{x}\to0$ describes the velocity distribution. In this case the corresponding quantities $\{\lambda:=\Lambda\left(\nabla_\mathbf{x}\to0\right)\}$ are eigenvalues of the Fokker-Planck operator $L_v$. In the collection $\{\lambda\}$ there is one zero eigenvalue denoted by $\lambda_0=0$ and the related eigenfunction is the stationary distribution of the velocity $\mathbf{v}$. All the other eigenvalues $\lambda$ have negative real parts, so that the corresponding eigenvectors describe decaying deviations from the stationary distribution. 

Inserting ansatz~\eqref{Fr1} into the Fokker-Planck equation~\eqref{Fr0} and separating terms with different eigenfunctions we obtain the governing equations for the coefficients $f_\Lambda(\mathbf{x},t)$
\begin{equation}\label{Fr3}
    \frac{\partial f_\Lambda(\mathbf{x},t)}{\partial t} = \Lambda(\nabla_\mathbf{x}) f_\Lambda(\mathbf{x},t)\,.
\end{equation}
The quantities $f_\Lambda(\mathbf{x},t)$ except for one matching the eigenvalue $\Lambda_0(\nabla_\mathbf{x})$ originating from $\lambda_0=0$ decay in time rather fast due to the fact that the corresponding eigenvalues $\lambda < 0$ of the zeroth approximation are negative and split from zero by a finite gap.  So the long time dynamics of the random variable $\mathbf{x}$ is governed by a generalized diffusion equation
\begin{equation}\label{Fr4}
    \frac{\partial f_0(\mathbf{x},t)}{\partial t} = \Lambda_0(\nabla_\mathbf{x}) f_0(\mathbf{x},t)\,.
\end{equation}
It is the standard diffusion equation if the eigenvalue $\Lambda_0(\nabla_\mathbf{x})$ is an analytic function of the operator $\nabla_\mathbf{x}$. When the functional  $\Lambda_0(\nabla_\mathbf{x})$ is non-analytic equation~\eqref{Fr4} describes anomalous diffusion. 

The purpose of the present work is to demonstrate that the non-analytic behavior of the eigenvalue $\Lambda_0(\nabla_\mathbf{x})$ becomes possible when the second moment of the stationary velocity distribution corresponding to the Fokker-Planck operator $L_v$ diverges.

\section{Velocity distribution}

It is the statistical properties of the particle velocity $\mathbf{v}$ that are responsible for the L\'evy type dynamics of the given particle. So the present section is devoted to their individual analysis. It should qualitatively explain the results to be obtained below and elucidate the mechanism via which the L\'evy type process arises.

The distribution of particle velocity is specified by the partial distribution function
\begin{equation}\label{FP3}
  P_{v}(\mathbf{v},\mathbf{v}_{0},t)=\int_{\mathbb{R}^{N}}d\mathbf{x}\,
  \mathcal{P}\left(  \mathbf{x}-\mathbf{x}_{0},\mathbf{v},\mathbf{v}_{0},t\right)
\end{equation}
obeying the reduced Fokker-Planck equation
\begin{equation}\label{FP4}
    \tau\frac{\partial P_{v}}{\partial t}=\sum_{i=1}^{N}\frac{\partial}{\partial v_{i}}
    \left[ g^{2}(\mathbf{v})\frac{\partial P_{v}}{\partial v_{i}}+(N+\alpha) v_{i}P_{v}\right]
\end{equation}
with the initial condition
\begin{equation}\label{FP5}
  P_{v}\left(  \mathbf{v},\mathbf{v},0\right)  =\delta(\mathbf{v}-\mathbf{v}_{0})\,.
\end{equation}
These expressions stem directly from equation~\eqref{FPKlim} and the initial condition~\eqref{FP2} after integrating them over $\mathbb{R}^N$.

Below in this section the velocity extremum distribution will be analyzed based on the first passage time problem. For this purpose we need another equation governing the velocity distribution $P_{v}(\mathbf{v},\mathbf{v}_{0},t)$ and acting on the initial velocity $\mathbf{v}_0$. Namely, it is the following backward Fokker-Planck equation of the \^Ito type
\begin{equation}\label{FP4b}
    \tau\frac{\partial P_{v}}{\partial t}=\sum_{i=1}^{N}
    \left[ g^{2}(\mathbf{v}_0)\frac{\partial^2 P_{v}}{\partial v^2_{0,i}}-(N+\alpha-2) v_{0,i}\frac{\partial  P_{v}}{\partial v_{0,i}}\right]
\end{equation}
conjugated with the forward Fokker-Planck equation~\eqref{FP4} of the H\"anggi-Klimontovich type and subjected to the same initial condition~\eqref{FP5} (see, e.g., Ref.~\cite{Gardiner}).

\subsection*{Stationary distribution}

Equation~\eqref{FP4} admits the stationary solution $P_{v}^{\text{st}}(\mathbf{v})$ obeying the condition of zero value probability flux
\begin{equation}\label{db}
  g^{2}(\mathbf{v})\frac{\partial P_{v}^{\text{st}}}{\partial v_{i}}
  +(N+\alpha) v_{i}P_{v}^{\text{st}}=0
\end{equation}
and the normalization to unity
\begin{equation}\label{norma}
   \int_{\mathbb{R}^{N}}d\mathbf{v}\,P_{v}^{\text{st}}(\mathbf{v})=1\,.
\end{equation}
Equation~\eqref{db} and condition~\eqref{norma} directly yield
\begin{equation}\label{VFPStat}
  P_{v}^{\text{st}}(\mathbf{v})=\frac{\Gamma\left[(N+\alpha)/2\right]}{\pi^{N/2}\Gamma\left( \alpha/2\right) }
  \,\frac{v_a^{\alpha}}{[g(\mathbf{v})]^{N+\alpha}}\,,
\end{equation}
where $\Gamma(\ldots)$ is the Gamma function. It should be noted that expression~\eqref{db} reflects the fact that the given random process admits the detailed balance with respect to the particle velocity treated individually. For the exponent $\alpha$ belonging to interval~\eqref{alpha} the first moment of the velocity magnitude $v=\left\vert \mathbf{v}\right\vert $ converges whereas the second one diverges.

The found expression~\eqref{VFPStat} for the velocity distribution actually indicates that the particle displacement $\delta \mathbf{r}\sim \mathbf{v}\delta t$ in the space $\mathbb{R}^N$ during some time interval $\delta t$ is likely to exhibit also similar power-law distribution typical for the L\'evy type random processes.

\subsection*{Moment dynamics}

The L\'evy flights and, partly, the L\'evy random walks are characterized by mutually independent succeeding steps in the particle displacement. This subsection analyzing the dynamics of velocity moments illustrates us that the proposed model does exhibit the loss of correlation in the particle velocity on time scales exceeding the value $\tau$. Thereby the partition of a particle trajectory into segments of duration $\delta t\gg\tau$ really can be regarded as a sequence of independent particle jumps.

In order to analyze the time dependence of the velocity moments, equation~\eqref{FP4} is multiplied, at first, with a general function $\omega(\mathbf{v})$ and integrated over all the possible values of the particle velocity. In this way we get the equality
\begin{equation*}
\tau\frac{d\left<\omega\right> }{dt}=
\left< g^2(\mathbf{v}) \sum_{i=1}^{N}\frac{\partial^2 \omega}{\partial v_i^2}\right>
-\left( \alpha+N-2\right) \left<  \sum_{i=1}^{N} v_{i}\frac{\partial\omega}{\partial v_{i}}\right>
\end{equation*}
where the symbol $\left<\ldots\right>$ means the standard averaging procedure, namely,
\begin{equation*}
\left< \ldots\right> =
 \int\limits_{\mathbb{R}^N} d\mathbf{v}\,(\ldots)P_v(\mathbf{v},\mathbf{v_0},t)\,.
\end{equation*}
Then setting $\omega(\mathbf{v})=v_{i}$, $\left\vert \mathbf{v}\right\vert $, $v_{i}v_{j}$ $(i\neq j)$, and $\mathbf{v}^{2}$ the obtained equation is converted into the governing equations for the corresponding velocity moments
\begin{align}
\label{mom1}
   \tau\frac{d\left< v_{i}\right>}{dt}  &  =-\left(N+\alpha-2\right) \left< v_{i}\right>\,,
\\
\label{mom2}
  \tau\frac{d\left<\left|\mathbf{v}\right|\right>}{dt}& = -\left(\alpha-1\right)\left< \left|\mathbf{v}\right|\right>
  + \mathcal{R}_N(t)\,,\\
\intertext{where}
\nonumber
   & \mathcal{R}_N(t) =
\begin{cases}
  2v_a^{2}P_{v}(0,t) & \text{for $N=1$,}\\[8pt]
 (N-1)v_a^{2}\left< \dfrac{1}{\left|\mathbf{v}\right|}\right>  & \text{for $N>1$,}
\end{cases}
\\
\label{mom3}
\frac{\tau}2\frac{d\left< v_{i}v_{j}\right>}{dt}  &  =-\left(N+\alpha-2\right)  \left< v_{i}v_{j}\right> \,,\quad\text{for}\quad i\neq j
\\
\label{mom4}
 \frac{\tau}2\frac{d\left<\mathbf{v}^{2}\right>}{dt}  &  =\left(2-\alpha\right)  \left<\mathbf{v}^{2}\right> + Nv_a^{2}\,.
\end{align}
According to equations~\eqref{mom1} and \eqref{mom3}, on time scales about $\tau$  the particle forgets the direction of its initial motion and the motion along different axes becomes independent. Equation~\eqref{mom2} demonstrates the fact that the first moment of the velocity $\mathbf{v}$ remains bounded during the system motion and attains its equilibrium value at the same time scales. This feature which will be used further in analyzing the spectral properties of equation~\eqref{FPKlim}. Finally, equation~\eqref{mom4} shows us that the system has to get its stationary state on temporal scales actually exceeding the parameter $\tau$ independently of the initial velocity.

\subsection*{Velocity maximum distribution}

Previously \cite{We} we claimed and justified numerically the fact that the L\'evy type behavior of the analyzed random walks in the 1D case on time scales $t\gg\tau$ is caused by the properties of extreme fluctuations in the particle velocity. Namely, an anomalously long displacement of the particle during time interval $t$ is approximately determined by its motion during the spike in the velocity fluctuations with the maximal amplitude. We will substantiate this statement also in the general case comparing the results of the present subsection devoted to maximum statistics of the given random walks and the distribution function of particle spatial displacements to obtained further.

Let us consider the first passage time problem for the model at hand. The probability $F(\mathbf{v}_{0},\vartheta ,t)$ for the particle with initial velocity $\mathbf{v}_{0}$ such that $\left\vert \mathbf{v}_{0}\right\vert <\vartheta $ to reach the sphere $|\mathbf{v}|= \vartheta$ in the velocity space for the first time at the instant $t$ is directly described by the backward Fokker-Planck equation~(\ref{FP4b}). In particular its Laplace transform
\begin{equation*}
  F_{L}(\mathbf{v}_{0},\vartheta ,s)=\int_{0}^{\infty }dt\,e^{-st}F(\mathbf{v}_{0},\vartheta ,t)
\end{equation*}
obeys the equation (see, e.g., \cite{Gardiner})
\begin{equation}\label{FPTP.1}
   \tau s F_L =\sum_{i=1}^{N}
    \left[ g^{2}(\mathbf{v}_0)\frac{\partial^2 F_L}{\partial v^2_{0,i}}-(N+\alpha-2) v_{0,i}\frac{\partial  F_L}{\partial v_{0,i}}\right]
\end{equation}
subject to the boundary condition
\begin{equation}\label{FPTP.2}
\left. F_{L}(\mathbf{v}_{0},\vartheta ,s)\right\vert _{\mathbf{v}_{0}= \vartheta }=1\,.
\end{equation}
Due to the symmetry of problem~\eqref{FPTP.1}, \eqref{FPTP.2} its solution $F_{L}(\mathbf{v}_{0},\vartheta ,s)$ is a symmetrical function of the argument $\mathbf{v}_0$ and can be treated as a function $F_{L}(v_{0},\vartheta ,s)$ of the magnitude
\begin{equation*}
 v_0:= \left(\sum_{i=1}^N v_{0,i}^2\right)^{1/2}\,.
\end{equation*}
In this terms equation~\eqref{FPTP.1} can be rewritten as
\begin{align}
\nonumber
   \tau s F_L = &
    g^{2}(v_0)\left[ \frac{\partial^2 F_L}{\partial v^2_{0}} + \frac{(N-1)}{v_0} \frac{\partial  F_L}{\partial v_{0}}\right]
\\
\label{FPTP.1s}
    &{}-(N+\alpha-2) v_{0}\frac{\partial  F_L}{\partial v_{0}}\,,
\end{align}
where the function $g^{2}(v_0)$ is given by formula~\eqref{4}.

The introduced first passage time probability is necessary to analyze the velocity maximum distribution. Namely we need the probability $\Phi \left(
\mathbf{v}_{0},\vartheta ,t\right) $ for the velocity pattern $\mathbf{v}(t)$ originating from the point $\mathbf{v}_{0}$ such that $\mathbf{v}_{0}<\vartheta$ to the get the maximum $|\mathbf{v}| = \vartheta$ during the time interval $t$ is related to the probability $F(\mathbf{v}_{0},\vartheta,t)$ by the following expression~\cite{extrema}
\begin{align}
\label{ed:2a}
    \Phi (\mathbf{v}_{0},\vartheta ,t)& =-\frac{\partial }{\partial \vartheta}\int_{0}^{t}dt^{\prime}\,F({v}_{0},\vartheta ,t^{\prime })
\\
\intertext{or for the Laplace transforms}
\label{ed:2b}
    \Phi _{L}(\mathbf{v}_{0},\vartheta,s)&=-\frac{1}{s}\frac{\partial }{\partial \vartheta }F_{L}({v}_{0},\vartheta,s)\,.
\end{align}
So the desired probability $\Phi (\mathbf{v}_{0},\vartheta ,t) = \Phi (v_{0},\vartheta ,t)$ as well as its Laplace transform actually depends on the magnitude $v_0$ of the vector $\mathbf{v}_0$ rather then on its components individually.

To examine the characteristic properties of the first passage time statistics let us consider the limit case $s\tau\ll1$ and $\vartheta\gg v_a$. Under these conditions in agreement with the results to be obtained we can regard the function $F_{L}(v_{0},\vartheta ,s)$ to be approximately constant  $F_{L}(v_{0},\vartheta ,s)\simeq F_{0}(\vartheta ,s)$ inside some neighborhood $\mathbb{Q}_0$ of the origin, $v_{0}=0$, whose thickness is much larger than $v_a$. In particular, for $s\rightarrow 0$ it is the sphere $|\mathbf{v}|< \vartheta$ itself and $F_{0}(\vartheta ,s)=1$ by virtue of \eqref{FPTP.2}. Within the neighborhood $\mathbb{Q}_0$ equation~\eqref{FPTP.1s} can be integrated directly with respect to the formal variable $f(v_{0}):=\partial F_{L}/\partial v_{0}$ using the standard parameter-variation method. In this way taking into account that $f(0)=0$ due to the system symmetry we obtain the expression
\begin{widetext}
\begin{align}
\label{FPTP.3}
    \frac{\partial F_{L}(v_{0},\vartheta ,s)}{\partial v_{0}} &\simeq \frac{\tau s}{v_{a}}F_{0}(\vartheta ,s)
    \left( \frac{v_{0}^{2}}{v_{a}^{2}}+1\right)^{\tfrac{N+\alpha -2}{2}}\left(\frac{v_a}{v_0}\right)^{N-1}
    \int\limits_{0}^{v_{0}/v_{a}}\frac{\xi^{N-1}\, d\xi}{\left(\xi ^{2}+1\right) ^{\tfrac{N+\alpha}{2}}}
\\
\intertext{and for $v_{0}\gg v_{a}$}
\label{FPTP.5}
    \frac{\partial F_{L}(v_{0},\vartheta ,s)}{\partial v_{0}} &\simeq \tau s\,F_{0}(\vartheta ,s)
    \frac{\Gamma \left(\frac{N}{2}\right)\Gamma \left(\frac{\alpha }{2}\right) }{2\Gamma \left( \frac{N+\alpha}{2}\right) }
    \frac{v_{0}^{\alpha -1}}{v_{a}^{\alpha }}\,.
\end{align}
\end{widetext}
Expression~\eqref{FPTP.5}, first, enables us to estimate the size $\bar{\vartheta}_L(s)$ of the domain $\mathbb{Q}_0$. In fact, inside the domain $\mathbb{Q}_0$ the inequality
\begin{equation*}
    F_0(\vartheta,s) \gg v_0 \frac{\partial F_{L}(v_{0},\vartheta ,s)}{\partial v_{0}}
    \quad\Rightarrow\quad
    \tau s\frac{v_{0}^{\alpha }}{v_{a}^{\alpha }}\ll 1
\end{equation*}
has to hold, which allows us to set
\begin{align}
\label{FPTP.4L}
    \bar{\vartheta}_L(s)& \sim \left( \frac1{\tau s}\right)^{\frac{1}{\alpha }}v_{a}\gg v_0
\intertext{or converting to the time dependence}
\label{FPTP.4}
    \bar{\vartheta}(t) & = \left( \frac{t}{\tau }\right)^{\frac{1}{\alpha }}v_{a}\gg v_0\,.
\end{align}
So the characteristic velocity scale characterizing the first passage time probability and aggregating its time dependence is $\bar{\vartheta}(t)$.

Second, in the case $\vartheta \gg \bar{\vartheta}_L(s)$  there is a spherical layer $v_{a}\ll v_{0} \ll \bar{\vartheta}(t)$ (for $N=1$ it is a couple of domains) wherein the equality $F_{L}(v_{0},\vartheta ,s)\simeq F_{0}(\vartheta ,s)$ holds whereas the derivative  $\partial F_{L}/\partial v_{0}$ scales with $v_{0}$ as $\partial F_{L}/\partial v_{0}\propto v_{0}^{\alpha -1}$. This asymptotic behavior can be obtained also analyzing the solution of equation~\eqref{FPTP.1} for $\left\vert v_{0}\right\vert \gg v_{a}$ where $g^{2}(v_{0})\simeq v_{0}^{2}$. In this case equation~\eqref{FPTP.1s} admits two solutions of the form
\begin{gather}
\nonumber
    F_{L}(v_{0},\vartheta ,s)\propto v_{0}^{g_{1,2}}
\\
\intertext{with}
\label{FPTP.6}
    g_{1}\simeq \alpha \quad \text{and}\quad g_{2}\simeq -\frac{\tau s}{\alpha }\,.
\end{gather}
The second solution is relevant to the function $F_{L}(v_{0},\vartheta ,s)$ only within the crossover from $F_{L}(v_{0},\vartheta ,s)\propto v_{0}^{\alpha}$ to $F_{L}(v_{0},\vartheta ,s)\approx $ $F_{0}(\vartheta ,s)$ and even in this region, i.e. $v_{0}\lesssim \bar{\vartheta}(t)$ the derivative $\partial F_{L}/\partial v_{0}$ is determined by its asymptotics $F_{L}(v_{0},\vartheta ,s)\propto v_{0}^{\alpha }$. For larger values of $v_0$, i.e., $v_{0} \gg \bar{\vartheta}(t)$ the first passage time distribution is given by the expression
\begin{equation}\label{AH.1}
    F_{L}(v_{0},\vartheta ,s)\simeq \left( \frac{v_{0}}{\vartheta}\right)^{\alpha}
\end{equation}
taking into account the boundary condition~\eqref{FPTP.2}. So we can write
\begin{equation}\label{FPTP.7}
    \frac{\partial F_{L}(v_{0},\vartheta ,s)}{\partial v_{0}}\simeq \alpha
    \frac{ v_{0}^{\alpha -1}}{\vartheta ^{\alpha }}
\end{equation}
also for $v_{0}\lesssim \bar{\vartheta}(t)$.

Expressions~\eqref{FPTP.5} and \eqref{FPTP.7} describe the same asymptotic behavior of the function $F_{L}(v_{0},\vartheta ,s)$. Thereby we can ``glue'' them together, obtaining the expression
\begin{equation}\label{FPTP.9}
    F_{0}(\vartheta ,s)=\frac{2\alpha \Gamma \left( \frac{N+\alpha}{2}\right) }
    {\Gamma \left( \frac{N }{2}\right)\Gamma \left( \frac{\alpha }{2}\right) }\,\frac{1}{\tau s}
    \left(\frac{v_{a}}{\vartheta}\right)^{\alpha }\,,
\end{equation}
that holds in the limit $\vartheta \gg \bar{\vartheta}_L(s)$. It should be noted that this procedure is the kernel of the singular perturbation technique which will be also used below.

Expression~\eqref{FPTP.9} immediately gives us the desired formula for the maximum distribution $\Phi _{L}(v_{0},\vartheta ,s)$. Namely, by virtue of
\eqref{ed:2b}, for $v_{0}\lesssim \bar{\vartheta}_L(s)$ and $\vartheta \gg \bar{\vartheta}_L(s)$ we have
\begin{equation}
\label{ed:5a}
    \Phi _{L}(v_{0},\vartheta ,s)=\frac{2\alpha ^{2}\Gamma\left( \frac{N+\alpha}{2}\right) }
    {\Gamma \left( \frac{N }{2}\right)\Gamma \left( \frac{\alpha }{2}\right) }\frac{1}{\tau s^{2}}
    \frac{v_{a}^{\alpha }}{\vartheta ^{\alpha +1}}\,.
\end{equation}
Then restoring the time dependence of the extremum distribution from its Laplace transform the asymptotic behavior for $\vartheta \gg \bar{\vartheta}(t)$ we get
\begin{equation}\label{ed:5b}
  \Phi (v_{0},\vartheta ,t)=\frac{2\alpha ^{2}\Gamma \left( \frac{N+\alpha}{2}
  \right) }{\Gamma \left( \frac{N }{2}\right)\Gamma \left( \frac{\alpha }{2}\right) }\frac{t}{\tau }
  \frac{v_{a}^{\alpha }}{\vartheta ^{\alpha +1}}\,.
\end{equation}
Finalizing the present subsection it is possible to draw the conclusion that for $v_{0}\ll\bar{\vartheta}(t)$ the maximum velocity distribution is described by a certain function
\begin{equation}\label{ad:6a}
    \Phi (v_{0},\vartheta ,t)=\frac{1}{\bar{\vartheta}(t)}\Phi _{0}
    \left( \frac{\vartheta }{\bar{\vartheta}(t)}\right)
\end{equation}
with the asymptotics
\begin{equation}\label{ad:6b}
     \Phi _{0}\left( \xi \right) =\frac{2\alpha ^{2}\Gamma \left( \frac{N+\alpha}{2}\right) }
     {\Gamma \left( \frac{N}{2}\right)\Gamma \left( \frac{\alpha }{2}\right) }\,\frac{1}{ \xi^{\alpha +1}}\,.
\end{equation}
Here the velocity scale $\bar{\vartheta}(t)$ is given by expression~\eqref{FPTP.4}. We remind that distribution~\eqref{ad:6a} describes
the magnitude of the velocity maximum attained during time interval $t$. For $v_0\ll \bar{\vartheta}(t)$ all the directions of particle motion are equivalent. So for a multidimensional space, $N>1$, the probability density of the maximum velocity attained during time interval $t$ being equal to $\boldsymbol{\vartheta}$ is given by the function
\begin{align}
\label{ppzN}
 \tilde{\Phi} (v_{0},\boldsymbol{\vartheta} ,t) & = \frac1{S_N} \Phi (v_{0},|\boldsymbol{\vartheta}|,t)\,,
\\
\intertext{where $S_N$ is the area of the sphere of radius $|\boldsymbol{\vartheta}|$ in $\mathbb{R}^N$}
\nonumber
 S_N & =\frac{2\pi^{N/2} |\boldsymbol{\vartheta}|^{N-1}}{\Gamma\left(\frac{N}2\right)}\,.
\end{align}
In 1D space this sphere is degenerated into two points $\pm\vartheta$, so the corresponding probability is
\begin{equation}\label{ppz1}
 \tilde{\Phi} (v_{0},\pm\vartheta ,t) = \frac12 \Phi (v_{0},\vartheta,t)
\end{equation}
because of the symmetry of the velocity fluctuations.

\section{Generating function\label{sec:genfun}}

\subsection*{General relations}

The present section formally substantiates that model~\eqref{2Klim1}, \eqref{2Klim2} does exhibit the L\'evy type behavior on time scales $t\gg\tau$. For this purpose let us analyze the properties of the generating function introduced as follows
\begin{equation}\label{GenF}
  \mathcal{G}(\boldsymbol{\varkappa},\mathbf{k},t):=\left\langle \exp\left\{\frac{\mathrm{i}}{v_{a}\tau}
  \left[  \left(\mathbf{x}-\mathbf{x}_0\right)\boldsymbol{\varkappa}+\tau\mathbf{vk}\right]  \right\}  \right\rangle\,,
\end{equation}
where the used averaging operator $\left<\ldots\right>$ is specified by the expression
\begin{equation}\label{averaging}
\left< \ldots\right> =
 \int\limits_{\mathbb{R}^N\times\mathbb{R}^N} d\mathbf{x}d\mathbf{v}\,(\ldots)\mathcal{P}(\mathbf{x}-\mathbf{x}_0,\mathbf{v},\mathbf{v_0},t)
\end{equation}
and the wave vectors $\boldsymbol{\varkappa}$ and $\mathbf{k}$ are dimensionless variables. Then by virtue of the Fokker-Planck equation~\eqref{FPKlim} and the initial condition~\eqref{FP2} the generating function obeys the governing equation
\begin{multline}\label{FPGenF}
    \tau\frac{\partial\mathcal{G}}{\partial t}  =\sum_{i=1}^{N}\left\{\frac{\partial}{\partial k_{i}}
     \left(  \mathbf{k}^{2}\frac{\partial\mathcal{G}}{\partial k_{i}}\right)\right.
\\
   {}  + \left.[\varkappa_{i}-(N+\alpha)k_i] \frac{\partial\mathcal{G}}{\partial k_{i}}\right\}
 -\mathbf{k}^{2}\mathcal{G}
\end{multline}
subject to the initial condition at $t=0$
\begin{equation}\label{GenFICond}
\mathcal{G}(\boldsymbol{\varkappa},\mathbf{k},0)=\exp\left\{ \frac{\mathrm{i}}{v_{a}}\mathbf{v}_{0}\mathbf{k}\right\}  \,.
\end{equation}
At the origin $\mathbf{k}=\mathbf{0}$ and $\boldsymbol{\varkappa}=\mathbf{0}$ the function~\eqref{GenF} meets also the condition
\begin{equation}\label{GenFNorm}
\mathcal{G}(\mathbf{0},\mathbf{0},t)=1
\end{equation}
which follows directly from the normalization of the distribution function to unity. In deriving equation~\eqref{FPGenF} the following correspondence between the operators acting in the spaces $\{\mathbf{x},\mathbf{v}\}$ and $\{\boldsymbol{\varkappa},\mathbf{k}\}$
\begin{align*}
  \frac{\partial}{\partial x_{i}}&\leftrightarrow -\frac{\mathrm{i}}{v_{a}\tau}\varkappa_{i}\,,
   &
  \frac{\partial}{\partial v_{i}}&\leftrightarrow -\frac{\mathrm{i}}{v_{a}}\varkappa_{i}\,,
   &
   v_{i}&\leftrightarrow -\mathrm{i}v_a\frac{\partial}{\partial k_{i}}
\end{align*}
as well as the commutation rule
\begin{equation*}
    \frac{\partial}{\partial k_{i}}k_{j}-k_{j}\frac{\partial}{\partial k_{i}}=\delta_{ij}
\end{equation*}
have been used.

The argument $\boldsymbol{\varkappa}$ enters equation~\eqref{FPGenF} as a parameter; the given equation does not contain any differential operator
acting upon the function $\mathcal{G}(\boldsymbol{\varkappa},\mathbf{k},t)$ via the argument $\boldsymbol{\varkappa}$. This feature enables us to raise a question about the spectrum properties of equation~\eqref{FPGenF}, where the variable $\boldsymbol{\varkappa}$ is treated as a parameter given beforehand. Then the desired eigenfunctions and their eigenvalues
\begin{equation}\label{eigenFV}
    \left\{\Psi_\Lambda\left(\mathbf{k}|\boldsymbol{\varkappa},\wp\right)\right\}\,,
    \qquad\left\{  \Lambda\left(  \boldsymbol{\varkappa},\wp\right)\right\}
\end{equation}
obey the equation
\begin{multline}\label{Eigen1}
  -\Lambda\Psi_{\Lambda}=
   \sum_{i=1}^{N}\left\{\frac{\partial}{\partial k_{i}}\left(\mathbf{k}^{2}\frac{\partial\Psi_{\Lambda}}{\partial k_{i}}\right) \right.
\\
 {}+\left.\left[\varkappa_i-(N+\alpha) k_i\right]\frac{\partial\Psi_\Lambda}{\partial k_i}\right\} -\mathbf{k}^2\Psi_\Lambda\,,
\end{multline}
where the symbol $\wp$ denotes the complete collection of the eigenfunction parameters for a fixed value of $\boldsymbol{\varkappa}$. We point out that the time dependence corresponding to these functions has been chosen in the form $\exp\{-\Lambda t/\tau\}$, which explains the minus sign on the left-hand side of \eqref{FPGenF} and dimensionless type of the eigenvalues $\left\{\Lambda\left(\boldsymbol{\varkappa},\wp\right)\right\}$.

In these terms the solution of equation~\eqref{FPGenF} can be written as the series
\begin{multline}\label{Series}
    \mathcal{G}(\boldsymbol{\varkappa},\mathbf{k},t|\mathbf{v}_0) = \sum_{\wp}f\left(\boldsymbol{\varkappa},\wp|\mathbf{v}_{0}\right)
    \Psi_{\Lambda}\left(\mathbf{k}|\boldsymbol{\varkappa},\wp\right)
\\
    {}\times\exp\left\{-\Lambda\left(\boldsymbol{\varkappa},\wp\right)\frac{t}{\tau}\right\} \,,
\end{multline}
where $\{f(\boldsymbol{\varkappa},\wp|\mathbf{x}_{0},\mathbf{v}_{0})\}$ are the coefficients of expansion~\eqref{Series} which meet the
equality
\begin{equation}\label{coeff}
  \sum_{\wp}f\left(\boldsymbol{\varkappa},\wp|\mathbf{v}_{0}\right)\Psi_{\Lambda}\left(\mathbf{k}|\boldsymbol{\varkappa},\wp\right)
  =\exp\left\{\frac{\mathrm{i}}{v_{a}}\mathbf{v}_{0}\mathbf{k}\right\}
\end{equation}
following from the initial condition~\eqref{GenFICond}. In agreement with the results to be obtained below the spectrum of the Fokker-Planck
equation~\eqref{FPGenF} is bounded from below by a nondegenerate minimal eigenvalue $\Lambda_{\text{min}}(\boldsymbol{\varkappa})\geq 0$, whereas the other eigenvalues are separated from it by a final gap. So, as time goes on and the inequality $t\gg\tau$ holds, the term corresponding to the minimal eigenvalue will be dominant and sum~\eqref{Series} is reduced to
\begin{equation}\label{Series1}
  \mathcal{G}(\boldsymbol{\varkappa},\mathbf{k},t) =
   f_{\text{min}}\left(\boldsymbol{\varkappa}|\mathbf{v}_{0}\right)
   \Psi_{\text{min}}\left(\mathbf{k}|\boldsymbol{\varkappa}\right)
   \exp\left\{  -\Lambda_{\text{min}}\left(\boldsymbol{\varkappa}\right)\frac{t}{\tau}\right\}
\end{equation}
on large time scales. Here $\Psi_{\text{min}}\left(\mathbf{k}|\boldsymbol{\varkappa}\right)$ is the eigenfunction matching the eigenvalue
$\Lambda_{\text{min}}$.

Whence several consequences follow. First, the general identity~\eqref{GenFNorm} holds at any moment of time, thereby
\begin{equation}\label{Series2}
\Lambda_{\text{min}}\left( \boldsymbol{0}\right) = 0\,.
\end{equation}
Second, on large time scales $t\gg\tau$ the system has to ``forget'' the initial velocity $\mathbf{v}_{0}$, so the expansion coefficient $f_{\text{min}}\left(\boldsymbol{\varkappa}\right) $ does not depend on $\mathbf{v}_{0}$ and it can be aggregated into the function $\Psi_{\text{min}}\left(  \mathbf{k}|\boldsymbol{\varkappa}\right)$. In this way the initial condition~\eqref{coeff} reads
\begin{multline}\label{coeffNew1}
  \Psi_{\text{min}}\left(\mathbf{k}|\boldsymbol{\varkappa}\right)+
  \sum_{\wp, \Lambda > \Lambda_\text{min}}
  f\left(\boldsymbol{\varkappa},\wp|\mathbf{v}_{0}\right)\Psi_{\Lambda}\left(\mathbf{k}|\boldsymbol{\varkappa},\wp\right)
\\
  =\exp\left\{\frac{\mathrm{i}}{v_{a}}\mathbf{v}_{0}\mathbf{k}\right\}
\end{multline}

The terms in sum~\eqref{coeffNew1} with $\Lambda > \Lambda_\text{min}$ determine the dependence of the generating function $\mathcal{G}(\boldsymbol{\varkappa},\mathbf{k},t|\mathbf{v}_0)$ on the initial velocity $\mathbf{v}_0$ and, thus, the corresponding coefficients $f\left(\boldsymbol{\varkappa},\wp|\mathbf{v}_{0}\right)$ must depend on $\mathbf{v}_0$. Finding the first derivative of both the sides of this equation with respect to $\mathbf{v}_0$ we have
\begin{multline}\label{coeffNew2}
   \sum_{\wp, \Lambda > \Lambda_\text{min}}
  \frac{\partial f\left(\boldsymbol{\varkappa},\wp|\mathbf{v}_{0}\right)}{\partial v_{0,i}}
   \Psi_{\Lambda}\left(\mathbf{k}|\boldsymbol{\varkappa},\wp\right)
\\
  =\frac{\mathrm{i}}{v_a}k_i\exp\left\{\frac{\mathrm{i}}{v_{a}}\mathbf{v}_{0}\mathbf{k}\right\}
\end{multline}
and for $\mathbf{k}=\mathbf{0}$ and \textit{any} value of the initial velocity $\mathbf{v}_0$
\begin{equation}\label{coeffNew3a}
   \sum_{\wp, \Lambda > \Lambda_\text{min}}
  \frac{\partial f\left(\boldsymbol{\varkappa},\wp|\mathbf{v}_{0}\right)}{\partial v_{0,i}}
   \Psi_{\Lambda}\left(\mathbf{0}|\boldsymbol{\varkappa},\wp\right)= 0\,.
\end{equation}
The latter feature, third, enables us to write individually
\begin{equation}\label{coeffNew3b}
   \Psi_{\Lambda}\left(\mathbf{0}|\boldsymbol{\varkappa},\wp\right)= 0\,,\quad\Lambda > \Lambda_\text{min}
\end{equation}
for all the eigenfunctions except for $\Psi_{\text{min}}\left(\mathbf{k}|\boldsymbol{\varkappa}\right)$. By virtue of \eqref{coeffNew1} and condition~\eqref{coeffNew3b} the eigenfunction $\Psi_{\text{min}}\left(\mathbf{k}|\boldsymbol{\varkappa}\right)$ meets the equality
\begin{equation}\label{PsiNormal}
   \Psi_{\text{min}}\left(\mathbf{0}|\boldsymbol{\varkappa}\right) = 1
\end{equation}
at $\mathbf{k} = \mathbf{0}$.

Summarizing the aforementioned we see that on large time scales $t\gg\tau$ the asymptotic behavior of the given generating function is the following
\begin{align}
\label{IamH1}
     \mathcal{G}(\boldsymbol{\varkappa},\mathbf{k},t)  &  =\Psi_{\text{min}}\left(\mathbf{k}|\boldsymbol{\varkappa}\right)
     \exp\left\{  -\Lambda_{\text{min}}\left(  \boldsymbol{\varkappa}\right)  \frac{t}{\tau}\right\}\,,
\\
\intertext{and due to \eqref{PsiNormal}}
\label{IamH2}
      \mathcal{G}(\boldsymbol{\varkappa},\mathbf{0},t)  &  =\exp\left\{-\Lambda_{\text{min}}\left(\boldsymbol{\varkappa}\right)
       \frac{t}{\tau}\right\}
\end{align}
for $\mathbf{k} = \mathbf{0}$. In what follows calculating the eigenvalue $\Lambda_{\text{min}}\left(\boldsymbol{\varkappa}\right)$ will be the main goal.

The randow walks under consideration should exhibit the L\'evy type behavior on large spatial and temporal scales, i.e. for $|\mathbf{x}-\mathbf{x}_0|\gg v_a\tau$ and $t\gg \tau$. This allows us to confine our analysis to the limit of small values of $\boldsymbol{\varkappa}$, i.e. to assume $\left\vert \boldsymbol{\varkappa }\right\vert \ll1$ and also $\Lambda_\text{min}(\boldsymbol{\varkappa})\ll 1$. Under such conditions the spectrum problem~\eqref{Eigen1} may be studies using perturbation technique with term
\begin{equation}\label{perturb}
\sum_{i=1}^{N}\varkappa_{i}\frac{\partial\Psi_{\Lambda}}{\partial k_{i}}
\end{equation}
playing the role of perturbation.

\subsection*{Zeroth approximation. Spectral properties of the velocity distribution}

The zeroth approximation of equation~\eqref{Eigen1} in perturbation~\eqref{perturb} matches the case $\boldsymbol{\varkappa} = \mathbf{0}$, where the generating function $\mathcal{G}(\mathbf{0},\mathbf{k},t)$ actually describes the velocity distribution as stems from its definition~\eqref{GenF}. Setting $\boldsymbol{\varkappa} = \mathbf{0}$ in the eigenvalue equation~\eqref{Eigen1} we reduce it to the following
\begin{multline}\label{Eigen0}
   -\lambda\Phi_{\lambda}=\sum_{i=1}^{N}\left\{\frac{\partial}{\partial k_{i}}\left(\mathbf{k}^{2}\frac{\partial\Phi_{\lambda}}
    {\partial k_{i}}\right)\right.
\\
  \left.{}-(N+\alpha) k_{i}\frac{\partial\Phi_{\lambda}}{\partial k_{i}}\right\}-\mathbf{k}^{2}\Phi_{\lambda}\,,
\end{multline}
where the designations
\begin{equation}\label{eigenFV0}
  \Phi_{\lambda}(\mathbf{k}|\wp)=\Psi_{\Lambda}\left(  \mathbf{k}|\mathbf{0},\wp\right)
  \quad\text{and}\quad
  \lambda\left(  \wp\right)  =\Lambda\left(\mathbf{0},\wp\right)
\end{equation}
have been used.

Pursuing different goals let us consider the conversion of equation~\eqref{Eigen0} under the replacement
\begin{equation}\label{evProb1}
  \Phi_{\lambda}\left(  \mathbf{k}|\wp\right)  =\left\vert \mathbf{k}\right\vert
  ^{\beta_{n}}\phi_{\lambda,n}\left(  \mathbf{k}|\wp\right)
\end{equation}
for several values of the exponent $\beta_{n}$.

\subsubsection*{Case 1:\quad $\beta_1 = (N+\alpha)/2$}

In this case the substitution of \eqref{evProb1} into equation~\eqref{Eigen0} converts it into
\begin{multline}\label{eqn1}
   \lambda\phi_{\lambda,1}=-\sum_{i=1}^{N}\frac{\partial}{\partial k_i}
    \left(\mathbf{k}^{2}\frac{\partial\phi_{\lambda,1}}{\partial k_{i}}\right)
\\
  {}  + \left[\mathbf{k}^{2}+ \frac{\alpha^2 -N^2}{4}\right]
    \phi_{\lambda,1}\,.
\end{multline}
The operator on the right-hand side of equation~\eqref{eqn1} is Hermitian within the standard definition of scalar product. So all the eigenvalues $\{\lambda\}$ are real numbers and the corresponding eigenfunctions $\{\phi_{\lambda,n}\left(  \mathbf{k}|\wp\right)\}$ form a basis. It should
be noted that the given conclusion coincides with the well known property of the Fokker-Planck equation for Markovian systems with the detailed balance \cite{Risken}. In addition the eigenfunctions $\phi_{\lambda,1}\left(\mathbf{k}|\wp\right)$ can be constructed in such a way that the identity
\begin{equation}
\int_{\mathbb{R}^{N}}d\mathbf{k}\,\phi_{\lambda,1}^{\ast}\left(
\mathbf{k}|\wp\right)  \phi_{\lambda^{\prime},1}\left(  \mathbf{k}|\wp
^{\prime}\right)  =\delta_{\wp\wp^{\prime}} \label{orthog}%
\end{equation}
hold for all of them except for the case describing the normalization of the eigenfunction $\phi_{0,1}\left(\mathbf{k}\right)$ corresponding to the minimal eigenvalue $\lambda_{\text{min}} = \Lambda_\text{min}(\mathbf{0})= 0$ by virtue of \eqref{Series2}. We note that the latter eigenfunction matches the stationary distribution~\eqref{VFPStat} of the particle velocity and its normalization is determined by condition~\eqref{PsiNormal}.

\subsubsection*{Case 2:\quad $\beta_2 = (N+\alpha-2)/2$}

In this case equation~\eqref{Eigen0} can be rewritten as
\begin{multline}\label{eqn2}
  \lambda\phi_{\lambda,2}= - \mathbf{k}^{2}\sum_{i=1}^N \frac{\partial^2 \phi_{\lambda,2}}{\partial k_i^{2}}
\\
   {}+ \left[\mathbf{k}^{2} + \frac{\alpha^2- (N-2)^2}{4}\right] \phi_{\lambda,2}\,.
\end{multline}
Let us split the Laplacian entering the right-hand side of expression~\eqref{eqn2} into two parts acting either on the magnitude $k:=\left\vert\mathbf{k}\right\vert$ of the vector $\mathbf{k}$ or on its angular variables
\begin{equation}\label{Loperator}
     \sum_{i=1}^N \frac{\partial^2}{\partial k_i^{2}} =
     \frac{1}{k^{N-1}}\frac{\partial}{\partial k}\left(k^{N-1}\frac{\partial}{\partial k}\right)
     -\frac{1}{k^{2}}\widehat{\mathbf{L}}^{2},
\end{equation}
where $\widehat{\mathbf{L}}$ is the angular momentum operator. We will remind the particular expressions of $\widehat{\mathbf{L}}^{2}$ for two- and
three-dimensional spaces
\begin{align*}
  \widehat{\mathbf{L}}_{N=2}^{2} & =-\frac{\partial^{2}}{\partial\varphi^{2}}\,,
\\
  \widehat{\mathbf{L}}_{N=3}^{2} &=-\frac{1}{\sin\theta}\frac{\partial}{\partial\theta}\left(\sin\theta\frac{\partial}{\partial\theta}\right)
 -\frac{1}{\sin^{2}\theta}\frac{\partial^{2}}{\partial\varphi^{2}}\,.
\end{align*}
In one-dimensional space the angular momentum operator $\widehat{\mathbf{L}}^{2}_{N=1}$ can be treated as the symmetry operator under the reflection $x\longmapsto-x$. All the eigenfunctions $\left\{\Theta_{\omega}\right\}$ of the operator $\widehat{\mathbf{L}}^{2}$ possess eigenvalues $\left\{  \delta\lambda_{\omega}\geq\delta\lambda_{0}\right\}  $ exceeding some positive number $\delta\lambda_{0}>0$ of order unity, naturally, except for the function $1_{\omega}$ not depending on the angular variables for which $\widehat{\mathbf{L}}^{2}1_{\omega}=0$.

This feature enables us to confine our further analysis to the eigenfunctions $\left\{  \Phi_{\lambda}\left(  k|\wp_{s}\right)  \right\}  $ depending only on the value $k=\left\vert \mathbf{k}\right\vert$. Indeed, expression~\eqref{Loperator} enables us to rewrite equation~\eqref{eqn2} as follows
\begin{multline}\label{eqn2sym}
  \lambda\phi_{\lambda,2} =
   - \frac{1}{k^{N-3}}\frac{\partial}{\partial k}\left(k^{N-1}\frac{\partial \phi_{\lambda,2}}{\partial k}\right)
\\
   {}+ \left[k^2 + \frac{\alpha^2- (N-2)^2}{4}\right] \phi_{\lambda,2}
   + \widehat{\mathbf{L}}^{2}\phi_{\lambda,2}\,.
\end{multline}
Thereby in the general case any eigenfunction $\Phi_{\lambda}\left(  \mathbf{k}|\wp\right)$ can be written as the product of the corresponding symmetrical eigenfunction and a certain eigenfunction of the angular momentum operator $\widehat{\mathbf{L}}^{2}$
\begin{equation*}
  \Phi_{\lambda}\left(  \mathbf{k}|\wp_{s} {\textstyle\bigcup}\omega\right)=\Phi_{\lambda}\left(  k|\wp_{s}\right)  \cdot\Theta_{\omega}\,,
\end{equation*}
and their eigenvalues are related as
\begin{equation*}
\lambda\left(\wp_{s}{\textstyle\bigcup}\omega\right)  =\lambda\left(  \wp_{s}\right)  +\delta\lambda_{\omega}\,.
\end{equation*}
So the given spectrum problem is splitted independently into its analysis with respect to the symmetrical eigenfunctions and the spectrum problem for the angular momentum operator.

\subsubsection*{Case 3:\quad $\beta_3 = \alpha/2$}

For the given value of the exponent $\beta$ and the symmetrical eigenfunctions $\phi_{\lambda,3}^{s}\left(k|\wp_{s}\right)$ equation~\eqref{Eigen0} is reduced to the modified Bessel differential equation
\begin{equation}\label{bessel}
  k^{2}\frac{d^{2}\phi_{\lambda,3}^{s}}{dk^{2}}+k\frac{d\phi_{\lambda,3}^{s}}{dk}
 -\left[k^{2}+\frac{\alpha^2}4-\lambda\right]  \phi_{\lambda,3}^{s}=0\,.
\end{equation}
Since the desired eigenfunctions should decrease as $k\rightarrow\infty$ the solution of equation~\eqref{bessel} is given by the modified Bessel function of the second kind
\begin{equation}\label{Knu}
    \phi_{\lambda,3}^{s}(k)\propto K_\nu(k)
\end{equation}
with the order $\nu=\sqrt{\alpha^{2}/4-\lambda}$ because
\begin{equation*}
   K_{\nu}(k)\sim\sqrt{\frac{\pi}{2k}}e^{-k}\quad\text{as}\quad k\rightarrow\infty\,.
\end{equation*}
for any value of the parameter $\nu$ \cite{specfun}.

Whence it follows that, first, there are no eigenfunctions with eigenvalues  $\lambda < 0$. Indeed, for small values of the argument $k$ the modified Bessel function $K_\nu(k)\propto k^{-\nu}$ when its order is a positive number $\nu > 0$, which is the case for $\lambda < \alpha^2/4$.  Under these conditions the trial function
\begin{equation*}
  \Phi(k):= k^{\beta_3} K_{\nu}(k)\propto k^{-\nu +\alpha/2}\quad\text{for}\quad k\ll 1
\end{equation*}
exhibits strong divergence with $k\to0$. Second, as it must, the value $\lambda_\text{min} =0$ is the minimal eigenvalue and the given trial function takes some finite value at $k=0$. The corresponding eigenfunction will be analyzed in detail below. Third, the interval $0<\lambda<\alpha^2/4$ does not contain any additional eigenvalue. In fact, otherwise, the trial function
\begin{equation*}
  \phi_{\lambda,1}(k)=\phi_{\lambda,3}(k)k^{\beta_{3}-\beta_{1}}\propto k^{-N/2}K_{\nu}(k)
\end{equation*}
would give rise to strong divergence in the normalization condition~\eqref{orthog}. Finally, the eigenvalues $\lambda>\alpha^{2}/4$ form the continuous spectrum of the given problem. The corresponding eigenfunctions via the modified Bessel function of the second kind with pure imaginary order exhibit strongly oscillatory behavior as $k\to 0$ and due to the preceding cofactor $k^{\beta_3}$ meet condition~\eqref{coeffNew3b}.

In order to construct the desired eigenfunction $\Phi_\text{min}(k)$ matching the minimal eigenvalue $\lambda_\text{min} =0$ let us make use of the expansion of the function $K_\nu(k)$ for small values of the argument $k$
\begin{equation}\label{Knuexp}
   K_{\nu}(k)=\frac{\Gamma(\nu)}{2^{1-\nu}k^{\nu}}\left[1-\left(\frac{k}{2}\right)^{2\nu}\frac{\Gamma(1-\nu)}{\Gamma(\nu+1)}+O(k^{2})\right]
\end{equation}
which is justified for the order $0<\nu<1$ (see, e.g., Ref.~\cite{specfun}). For $\lambda =0$ the order $\nu = \alpha/2$ and the latter inequality holds due to the adopted assumption~\eqref{alpha} about the parameter $\alpha$. Expression~\eqref{evProb1} and the obtained result~\eqref{Knu} specify the dependence of the eigenfunction $\Phi_\text{min}(k)\propto k^{\alpha/2} K_{\alpha/2}(k)$ on the argument $k$. In the case under consideration the general condition~\eqref{PsiNormal} reading $\Phi_\text{min}(0)=0$ together with asymptotics~\eqref{Knuexp} enables us to find the preceding constant. In this way the desired expression
\begin{multline}\label{VDzeroeigen}
  \Phi_{\text{min}}(\mathbf{k}) = \frac{2^{\frac{2-\alpha}{2}}}{\Gamma\left(\frac{\alpha}{2}\right)}
   k^{\frac{\alpha}{2}}K_{\frac{\alpha}{2}}(k)
\\
 =1-\left(\frac{k}{2}\right)^{\alpha}\frac{\Gamma\left(\frac{2-\alpha}{2}\right)}{\Gamma\left(\frac{2+\alpha}{2}\right)}+O(k^{2})~.
\end{multline}
is got. Expression~\eqref{VDzeroeigen} finalizes the analysis of the zeroth approximation.

Summarizing the aforementioned we draw the conclusion that at $\boldsymbol{\varkappa} = \mathbf{0}$ the spectrum of the Fokker-Planck equation~\eqref{FPGenF} for the generating function~\eqref{GenF} does contain zero eigenvalue $\Lambda _{\text{min}}\left( 0\right) =0$ corresponding to eigenfunction~\eqref{VDzeroeigen} which is separated from higher eigenvalues by a gap equal to $\alpha ^{2}/4$ (in units of $\tau $). We note that the given statement is in agreement with the conclusion about the spectrum properties for a similar stochastic process with multiplicative noise~\cite{MN1,MN2,MN3}.

\subsection*{The eigenvalue $\Lambda_{\text{min}}(\boldsymbol{\varkappa})$ for $\left\vert\boldsymbol{\varkappa}\right\vert \ll1$. Singular perturbation technique}

On time scales $t\gg\tau$ only rather small values of the wave vector $\boldsymbol{\varkappa}$ contribute substantially to the distribution of the random variable $\mathbf{x}$, Indeed, according to \eqref{IamH1} or \eqref{IamH2} only the wave vectors $\boldsymbol{\varkappa}$ that meet the inequality
\begin{equation*}
 \Lambda_\text{min}(\boldsymbol{\varkappa})\frac{t}{\tau} \lesssim 1
\end{equation*}
are essential. Due to $\Lambda_\text{min}(\boldsymbol{\varkappa})\to 0$ as $\boldsymbol{\varkappa}\to \mathbf{0}$ (see \eqref{Series2}) and the estimate $\Lambda_\text{min}\sim 1$ for $|\boldsymbol{\varkappa}|\sim 1$ this inequality is converted into $|\boldsymbol{\varkappa}|\lesssim \varkappa_t\ll1$ with $\varkappa_t\to0$ as $t\to\infty$. So in what follows the inequality $|\boldsymbol{\varkappa}|\ll 1$ will be assumed to hold beforehand.

The perturbation term~\eqref{perturb} mixes the eigenfunctions of zeroth approximation affecting the spectrum property of the generating function. However in the case under consideration perturbation~\eqref{perturb} disturbs the eigenfunction $\Phi_\text{min}(\mathbf{k})$ with $\lambda_\text{min}=0$ substantially only in a certain small neighborhood $\mathbb{Q}_\varkappa$ of the origin $\mathbf{k}=\mathbf{0}$ wherein $|\mathbf{k}|\lesssim |\boldsymbol{\varkappa}|$. It stems directly from the form of the eigenfunction equation~\eqref{Eigen1}. Outside the neighborhood $\mathbb{Q}_{\varkappa}$ the eigenvalue function $\Psi_{\text{min}}(\mathbf{k}|\boldsymbol{\varkappa})$ should practically coincide with
$\Phi_{\text{min}}(\mathbf{k})$.

\begin{figure}
\begin{center}
\includegraphics[width=80mm]{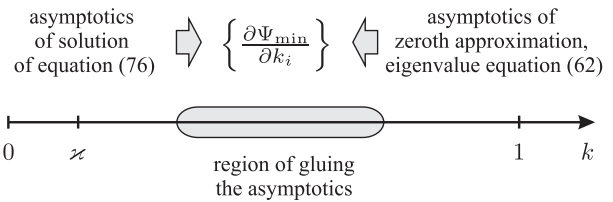}
\end{center}
\caption{Illustration of the singular perturbation technique in gluing the asymptotics the eigenfunction $\Psi_\text{min}(k)$ constructed in different limit region of the velocity wave number $\mathbf{k}$.}
\label{Fig1}
\end{figure}

So in the space of the wave numbers $\boldsymbol{\varkappa}$ there is a spherical layer $\mathbb{L}_\varkappa$
\begin{equation}\label{spt1}
\left\vert \boldsymbol{\varkappa}\right\vert \,\ll\left\vert \mathbf{k}\right\vert \ll 1
\end{equation}
wherein formula~\eqref{VDzeroeigen} in the limit $k\ll1$ approximates not only the eigenfunction $\Phi_\text{min}(\mathbf{k})$ but also the eigenfunction $\Psi_{\text{min}}(\mathbf{k}|\boldsymbol{\varkappa})$ (Fig.~\ref{Fig1}). In addition within this layer $\mathbb{L}_\varkappa$ as well as the neighborhood $\mathbb{Q}_{\varkappa}$, first, the the two eigenvalue functions are practically equal to unity, $\Psi_{\text{min}}(\mathbf{k}|\boldsymbol{\varkappa})$, $\Phi_\text{min}(\mathbf{k})\simeq1$. Second, any power term whose exponent exceeds
$\alpha < 2$ (see inequality~\eqref{alpha}) may be ignored. The latter concerns also the last term on the right-hand side of equation~\eqref{Eigen1}.

Keeping the aforesaid in mind and splitting the function $\Psi_{\text{min}}(\mathbf{k}|\boldsymbol{\varkappa})$ into two parts
\begin{equation}\label{smallpsi}
\Psi_{\text{min}}(  \mathbf{k}|\boldsymbol{\varkappa})=1-\psi(\mathbf{k}|\boldsymbol{\varkappa})\,,
\end{equation}
where $\psi(\mathbf{k}|\boldsymbol{\varkappa})\ll1$ the eigenvalue problem under consideration is reduced to solving the equation
\begin{equation}\label{spt2}
  \Lambda_{\text{min}}=\sum_{i=1}^{N}\left\{\frac{\partial}{\partial k_{i}}\left(\mathbf{k}^{2}\frac{\partial\psi}{\partial k_{i}}\right)
  +[\varkappa_{i}-(N+\alpha) k_{i}]  \frac{\partial\psi }{\partial k_{i}}\right\}
\end{equation}
subject to the requirement of the solution exhibiting the asymptotic behavior
\begin{equation}\label{spt3}
    \psi\left(\mathbf{k}|\boldsymbol{\varkappa}\right)\sim\left(\frac{k}{2}\right)^{\alpha}\frac{\Gamma\left(\frac{2-\alpha}{2}\right)}
   {\Gamma\left(\frac{2+\alpha}{2}\right)}\,.
\end{equation}
for $|\boldsymbol{\varkappa}| \ll |\mathbf{k}| \ll  1$. In some sense expression~\eqref{spt3} is a ``boundary condition'' of equation~\eqref{spt2}.

\subsubsection*{Scaling relations}

Using the scaling transformations
\begin{align}\label{scal1}
   k_{i} & =|\boldsymbol{\varkappa}|\,\zeta_{i},
&
  \Lambda_{\text{min}} & = |\boldsymbol{\varkappa}|^{\alpha}\varLambda_{\text{min}},
&
  \psi(\mathbf{k}|\boldsymbol{\varkappa}) & \mapsto|\boldsymbol{\varkappa}|^{\alpha}\psi(\boldsymbol{\zeta}|\mathbf{n})
\end{align}
the eigenvalue problem~\eqref{spt2}, \eqref{spt3} is reduced to the equation
\begin{equation}\label{scal2}
 \varLambda_{\text{min}}=\sum_{i=1}^{N}\left\{\frac{\partial}{\partial\zeta_{i}}
 \left(\zeta^{2}\frac{\partial\psi}{\partial\zeta_{i}}\right)  + \left[n_{i}-(N+\alpha)\zeta_{i}\right]
 \frac{\partial\psi}{\partial\zeta_{i}}\right\},
\end{equation}
where $\mathbf{n}=\{n_{i}\}$ is the unit vector parallel to the wave vector $\boldsymbol{\varkappa}$ and the asymptotic behavior~\eqref{spt3} is converted into the asymptotics
\begin{equation}\label{scal3}
  \psi\left(\boldsymbol{\zeta}|\mathbf{n}\right)
  \sim\left(\frac{|\boldsymbol{\zeta}|}{2}\right)^{\alpha}\frac{\Gamma\left(\frac{2-\alpha}{2}\right)}{\Gamma\left(\frac{2+\alpha}{2}\right)}
   \quad\text{as}\quad |\boldsymbol{\zeta}|\rightarrow\infty\,.
\end{equation}
at infinity. In some sense condition~\eqref{scal3} ``glues'' the asymptotic behavior of the eigenfunction $\Psi_{\text{min}}(\mathbf{k}|\boldsymbol{\varkappa})$ resulting from its properties for sufficiently large values of $\mathbf{k}$ together with one stemming from small values of $\mathbf{k}$, in this case, specified by the solution of equation~\eqref{scal2}. Exactly such a procedure is the essence of the singular perturbation technique.

The eigenvalue problem~\eqref{scal2}, \eqref{scal3} does not contain any external parameter, enabling us to expect that $\varLambda_{\text{min}}$ is a value about unity, $\varLambda_{\text{min}}\sim1$. This estimate actually is the basic result of the present paper because it immediately leads us to the conclusion about the scaling properties of the random walks in the space $\mathbb{R}^{N}$ that are governed by the system of equations~\eqref{2Klim1}) and \eqref{2Klim2}.

However before discussing the obtained results let us find the specific values of the eigenvalue $\varLambda_{\text{min}}$, with one-dimensional and multi-dimensional spaces being analyzed individually.

\subsubsection*{One-dimensional model}

For the one-dimensional space $\mathbb{R}$ ($N=1$) equation~\eqref{scal2} becomes
\begin{equation}\label{spt2b}
  \varLambda_{\text{min}}= \frac{d}{d\zeta}\left(\zeta^{2}\frac{d\psi}{d\zeta}\right)  + \left[1-(1+\alpha)\zeta\right]
 \frac{d\psi}{d\zeta}\,.
\end{equation}
Here we have assumed the unit vector $\mathbf{n}$ to be positively directed along the $\zeta$-axis and omitted the index $i$ at the variable $\zeta$.

Equation~\eqref{spt2b} can be solved directly with respect to the variable $d\psi/d\zeta$ using the standard parameter-variation method. However this solution can exhibit a strong singularity at $\zeta = 0$, so the regions $\zeta > 0$ and $\zeta <0$ have to be considered separately. In this way we
get for $\zeta <0$
\begin{multline}\label{1D:2a}
  \frac{d\psi}{d\zeta}  = |\zeta|^{\alpha -1}\exp \left(\frac1{\zeta} \right)
\\
  {}\times\left[C_{-\infty }+ \varLambda_\text{min}\int\limits_{0}^{-1/\zeta }\xi^{\alpha-1}e^{-\xi} d\xi \right]
\end{multline}
and for $\zeta>0$
\begin{multline}\label{1D:2b}
   \frac{d\psi }{d\zeta } =|\zeta|^{\alpha -1}\exp\left(\frac{1}{\zeta }\right)
\\
  {}\times \left[C_{+\infty } - \varLambda_\text{min} \int\limits_{0}^{1/\zeta }\xi^{\alpha-1}e^{-\xi} d\xi \right] \,,
\end{multline}
where the constants $C_{\pm \infty }$ specify the asymptotic behavior of the derivative
\begin{equation*}
  \frac{d\psi }{d\zeta }\sim \left\vert \zeta \right\vert ^{\alpha -1}C_{\pm\infty }
   \quad \text{as}\quad \zeta \rightarrow \pm \infty\,.
\end{equation*}
Thereby according to condition~\eqref{scal3} that should be imposed on the asymptotic behavior of the solution $\psi(\zeta)$ and converting into the corresponding asymptotic behavior of its derivative
\begin{equation}\label{1D:2c}
  C_{+\infty } = -C_{-\infty }=
  \frac{\alpha \Gamma \left( \frac{2-\alpha }{2} \right) }{2^{\alpha}\Gamma\left( \frac{2+\alpha }{2}\right) }\,.
\end{equation}
Solution~\eqref{1D:2b} diverges as $\zeta \rightarrow 0$ unless the equality
\begin{equation*}
  C_{+\infty}-\varLambda_\text{min}\int\limits_{0}^{\infty } \xi ^{\alpha -1}e^{-\xi}d\xi =0
\end{equation*}
holds, whence taking into account \eqref{1D:2c} we find the desired expression for the eigenvalue $\varLambda_{\text{min}}$
\begin{equation}\label{1D:3new}
  \varLambda _{\text{min}}=\frac{\Gamma\left(\frac{2-\alpha}{2}\right)}{2^{\alpha-1}\Gamma\left(\frac{\alpha }{2}\right)\Gamma\left(\alpha\right)}\,.
\end{equation}
In obtaining \eqref{1D:3new} the recurrence formulas for the Gamma function have been used.

\subsubsection*{Multi-dimensional model}

For the multi-dimensional space $\mathbb{R}^N$ ($N>1$) the solution of equation~\eqref{scal2} subject to the asymptotic behavior~\eqref{scal3} turns out to depend only on two variables, the absolute value $\zeta : = |\boldsymbol{\zeta}|$ of the wave vector $\boldsymbol{\zeta}$ and the angular variable
\begin{equation}\label{ND:0}
    \theta:=\sum_{i=1}^{N}\frac{\zeta_{i}n_{i}}{\zeta}\,.
\end{equation}
This statement follows directly from the symmetry of equation~\eqref{scal2} and the ``boundary condition''~\eqref{scal3}. Using the variables $\{\zeta,\theta\}$ we can write
\begin{align*}
 \frac{\partial}{\partial\zeta_i} & = \frac{\zeta_i}{\zeta}\frac{\partial}{\partial \zeta} +
 \frac{1}{\zeta^2}(\zeta n_i - \theta \zeta_i)\frac{\partial}{\partial \theta}\,,
\\
\intertext{and, thus,}
 \sum_{i=1}^N \zeta_i\frac{\partial}{\partial\zeta_i} & = \zeta\frac{\partial}{\partial \zeta}\,,
\\
 \sum_{i=1}^N n_i\frac{\partial}{\partial\zeta_i} & = \theta\frac{\partial}{\partial \zeta} +
 \frac{(1-\theta^2)}{\zeta}\frac{\partial}{\partial\theta}\,.
\end{align*}
assuming the given operators to act on functions of the arguments $\zeta$ and $\theta$ only. Then equation~\eqref{scal2} can be rewritten as
\begin{multline}\label{ND:1}
   \varLambda_{\text{min}}  = \zeta^{2}\frac{\partial^{2}\psi}{\partial\zeta^{2}}- (\alpha-1)\zeta\frac{\partial\psi}{\partial\zeta}
\\
   {}+ \left(  1-\theta^{2}\right)  \frac{\partial^{2}\psi}{\partial\theta^{2}}-\left(N-1\right)\theta
   \frac{\partial\psi}{\partial\theta}
\\
  {}+\frac{1}{\zeta}\left(1-\theta^{2}\right)\frac{\partial\psi}{\partial\theta}  +\theta\frac{\partial\psi}{\partial\zeta}\,.
\end{multline}
Thereby in the given case the eigenvalue problem is reduced to solving equation~\eqref{ND:1} subject to the boundary condition~\eqref{scal3} with respect to the function $\psi(\zeta,\theta)$.

The boundary $\theta = \pm 1$ is artificial; it arises via the change of variables~\eqref{ND:0}. So the solution should be an analytical function of the variable $\theta$ also all the points of the closed interval $\theta\in[-1,1]$. The latter feature enables us to seek the solution of equation~\eqref{ND:1} as a power series with respect to $\theta$
\begin{equation}\label{pwser.1}
     \psi(\zeta,\theta) = \sum_{n=0}^\infty \theta^n\psi_n(\zeta)\,.
\end{equation}
Substituting~\eqref{pwser.1} into \eqref{ND:1} and gathering terms with the same power of $\theta$ we obtain the collection of individual equations for the components of expansion~\eqref{pwser.1}, namely,
\begin{subequations}\label{pwser.2}
\begin{equation}\label{pwser.2a}
 \zeta^2\frac{d^2\psi_0}{d\zeta^2} - (\alpha-1) \zeta\frac{d\psi_0}{d\zeta}
  = -2 \psi_{2} - \frac{1}{\zeta}\psi_{1} + \varLambda_\text{min}
\end{equation}
and for $n\geq1$
\begin{multline}\label{pwser.2c}
 \zeta^2\frac{d^2\psi_n}{d\zeta^2} - (\alpha-1) \zeta\frac{d\psi_n}{d\zeta}- n(N+n-2)\psi_n
\\
{}= -(n+2)(n+1) \psi_{n+2} - \frac{(n+1)}{\zeta}\psi_{n+1}
\\
{} + \frac{(n-1)}{\zeta}\psi_{n-1}-\frac{d\psi_{n-1}}{d\zeta} \,.
\end{multline}
\end{subequations}
Let us, first, obtain the particular expression for the eigenvalue $\varLambda_\text{min}$ in the limit case $N\gg1$. According to equation~\eqref{pwser.2c} at the leading order in the small parameter $1/N$
\begin{equation}\label{largeN.1}
  \psi_{1}(\zeta) = \frac1N\cdot\frac{d\psi_0(\zeta)}{d\zeta}\,,
\end{equation}
whereas the other components are of higher order in $1/N$. So within the given accuracy equation~\eqref{pwser.2a} becomes.
\begin{equation}\label{largeN.2}
 \zeta^2\frac{d^2\psi_0}{d\zeta^2} +\left[\frac{1}{N\zeta} - (\alpha-1)\zeta\right] \frac{d\psi_0}{d\zeta}
  =  \varLambda_\text{min}\,.
\end{equation}
Then imposing the ``boundary'' condition~\eqref{scal3} on equation~\eqref{largeN.2} and using the method of variation of parameters it is solved, giving us the expression
\begin{multline}\label{largenN.3}
\frac{d\psi_0(\zeta)}{d\zeta} = \zeta^{\alpha-1}\exp\left\{\frac{1}{2N\zeta^2}\right\}
\\
{}\times\left[
\frac{\Gamma\left(\frac{2-\alpha}{2}\right)}{2^{\alpha-1}\Gamma\left(\frac{\alpha}{2}\right)}
-  2^{\tfrac{\alpha-2}{2}} N^{\tfrac{\alpha}2} \varLambda_\text{min}
 \int\limits_{0}^{\tfrac{1}{2N\zeta^2}}\xi^{\tfrac{\alpha}2-1}e^{-\xi}d\xi
\right]
\,.
\end{multline}
Therefor the function $\psi_0(\zeta)$ does not exhibit a singularity as $\zeta\rightarrow0$ if
\begin{equation}\label{ND:6}
 \varLambda_{\text{min}}=\left(\frac{1}{N}\right)^{\tfrac{\alpha}{2}}
      \frac{4\Gamma\left(\frac{2-\alpha}{2}\right)}{2^{\tfrac{3\alpha}{2}}\Gamma^{2}\left(\frac{\alpha}2\right)},
\end{equation}
which gives us the desired estimate of the eigenvalue $\varLambda_{\text{min}}$ at the leading order in $1/N$.

For an arbitrary value of $N$ the system~\eqref{pwser.2} could be analyzed numerically using, for example, the following algorithm. As follows from equations~\eqref{pwser.2}, in a small neighborhood of the origin $\zeta = 0$ the functions $\{\psi_n(\zeta)\}$ can be expanded into power series
\begin{equation*}
 \psi_n(\zeta) = \zeta^n\cdot\left[A_{n0} + A_{n1} \zeta^2 + A_{n2}\zeta^4 \ldots\right]\,,
\end{equation*}
where the coefficients $\{A_{nm}\}$ can be found by solving the system~\eqref{pwser.2}. In particular, beforehand, it is possible to set $A_{00} = 0$ due to the general equality~\eqref{PsiNormal} and definition~\eqref{smallpsi} of the function $\psi(\zeta,\theta)$. In addition, the equality $A_{10} = \varLambda_\text{min}$ stems directly from equation~\eqref{pwser.2a}. Therefore the given system of equations can be subjected to the following ``initial conditions''
\begin{align*}
   \left.\psi_{n}\right|_{\zeta = 0} &= 0\, &&\text{for all the values of the index $n$,}\\
   \left.\frac{d\psi_{n}}{d\zeta}\right|_{\zeta = 0} &= 0\, &&\text{for $n\neq1$,}\\	
   \left.\frac{d\psi_{1}}{d\zeta}\right|_{\zeta = 0} &= \varLambda_\text{min}&&\text{for $n=1$.}
\end{align*}
In this way the system of equations~\eqref{pwser.2a} and \eqref{pwser.2c} is converted into an initial value problem for any given value of the parameter $\varLambda_\text{min}$. The solution is proportional to the quantity $\varLambda_\text{min} = 1$ so, af first, setting, for example, $\varLambda_\text{min} = 1$ and solving this system numerically one finds a coefficient $\omega^*$ of the asymptotics $\psi_{0}(\zeta)\sim \omega^*\zeta^{\alpha}$ as $\zeta\to\infty$. Then
\begin{equation*}
  \varLambda_\text{min} =
   \frac{\Gamma\left(\frac{2-\alpha}{2}\right)}{\omega^*2^\alpha\Gamma\left(\frac{2+\alpha}{2}\right)}\,.
\end{equation*}
is the desired eigenvalue.

\section{L\'evy type distribution of spatial steps}

Summarizing the results of the previous section we draw the conclusion that the continuous time model~\eqref{2Klim1}, \eqref{2Klim2} for the random walks governed by the given multiplicative noise does exhibit the Lev\'{y} type behavior on time scales exceeding essentially the parameter $\tau$.

To justify this statement in detail, we take into account the scaling relations~\eqref{scal1} and rewrite the generating function~\eqref{IamH2} for the particle displacements in the space $\{\mathbf{x}\}$ in the form
\begin{equation}\label{final.1}
    \mathcal{G}(\boldsymbol{\varkappa},\mathbf{0},t)   =
    \exp\left\{-\varLambda_{\text{min}}\left|\boldsymbol{\varkappa}\right|^{\alpha}
    \frac{t}{\tau}\right\},
\end{equation}
which holds for $t\gg\tau$. Introducing the distribution function $P_{x}(\mathbf{x} - \mathbf{x}_{0},t)$ of the particle displacements by the integral
\begin{equation}\label{final.2}
  P_{x}(\mathbf{x} -\mathbf{x}_{0},t)= \int_{\mathbb{R}^{N}}d\mathbf{v}\,
  \mathcal{P}\left(\mathbf{x}-\mathbf{x}_{0},\mathbf{v},\mathbf{v}_{0},t\right)
\end{equation}
the generating function~\eqref{final.1} can be represented as
\begin{equation}\label{final.3}
    \mathcal{G}(\boldsymbol{\varkappa},\mathbf{0},t)   = \int\limits_{\mathbb{R}^N} d\mathbf{x}\, P_{x}(\mathbf{x} - \mathbf{x}_{0},t)
   \exp\left\{\frac{\mathrm{i}}{v_{a}\tau}(\mathbf{x}-\mathbf{x}_0)\cdot\boldsymbol{\varkappa} \right\} \,.
\end{equation}
Comparing expressions~\eqref{final.1} and \eqref{final.3} we see that, first, the distribution function $P_{x}(\mathbf{x} - \mathbf{x}_{0},t) = P_{x}(\delta x,t)$ is symmetrical, i.e. it should depend only on the length of the displacement vector $\delta x := |\mathbf{x} - \mathbf{x}_{0}|$.

Second, it does not depend on the initial velocity $\mathbf{v}_0$ of the particle, so on such time scales the description of these random walks can be confined to the spatial variables only.

Third feature is the time dependence of the spatial scale $\ell(t)$ characterizing the particle displacements during the time interval $t$. To find it we assume the argument of the generating function~\eqref{final.1} to meet the estimate $|\boldsymbol{\varkappa}|^\alpha (t/\tau)\sim 1$ for values of the wave vector $\boldsymbol{\varkappa}$ such that $(\ell(t)|\boldsymbol{\varkappa}|)/(v_a\tau)\sim 1$, see formula~\eqref{final.3}. When it immediately follows that
\begin{equation}\label{ell.def}
 \ell(t) = ( v_a \tau) \left( \frac{t}{\tau}\right)^{\tfrac{1}{\alpha}}\,.
\end{equation}
Due to the adopted inequality~\eqref{alpha} the exponent $1/\alpha> 1/2$, so the time dependence of the characteristic particle displacement $\ell(t)$ really describes a L\'evy type stochastic process.

Fourth characteristics is the asymptotic behavior exhibited by the distribution $P_{x}(\delta x,t)$ of the particle displacements. This asymptotics matches spatial scales $\delta x\gg\ell(t)$ and, correspondingly, small values of the argument of the generating function~\eqref{final.1}. Under these conditions expression~\eqref{final.3} is reduced to the following
\begin{equation}\label{final.4}
  \varLambda_\text{min}|\boldsymbol{\varkappa}|^{\alpha}\frac{t}{\tau} = 2\int\limits_{\mathbb{R}^N}d\mathbf{x}\,P_{x}(\delta x,t)
  \sin^2\left[\frac{(\mathbf{x}-\mathbf{x}_0)\cdot\boldsymbol{\varkappa}}{2v_a\tau}\right]
\end{equation}
obtained by subtracting the expression combined \eqref{final.1} and \eqref{final.3} from itself calculated at $\boldsymbol{\varkappa} = \mathbf{0}$.
In agreement with the results to be found below, the distribution function $P_{x}(\delta x,t)$ on scales $\delta x\gg\ell(t)$ is to exhibit the following asymptotic behavior
\begin{equation}\label{final.5}
   P_{x}(\delta x,t) \sim C\frac{[\ell(t)]^\alpha}{(\delta x)^{N+\alpha}}\,,
\end{equation}
where the constant $C$ can be found using expression~\eqref{final.4}. Namely, asymptotics~\eqref{final.5} enables us to rewrite expression~\eqref{final.4} for the one-dimensional case ($N=1$) as
\begin{subequations} \label{final.5N1}
\begin{equation}\label{final.51}
  \varLambda_\text{min}|\boldsymbol{\varkappa}|^{\alpha}\frac{t}{\tau} = 4C[\ell(t) ]^\alpha \int\limits_0^\infty
  \frac{d(\delta x)}{(\delta x)^{1+\alpha}}
    \sin^2\left[\frac{(\delta x)|\varkappa|}{2v_a\tau}\right]
\end{equation}
and for the multidimensional case ($N>1$) as
\begin{multline}\label{final.5N}
  \varLambda_\text{min}|\boldsymbol{\varkappa}|^{\alpha}\frac{t}{\tau} = 2C[\ell(t) ]^\alpha
   \int\limits_0^\pi d\varTheta \int\limits_0^\infty d(\delta x)
\\
  {}\times\frac{\delta x}{(\delta x)^{N+\alpha}}S_{N-1}(\delta x \sin\varTheta)
    \sin^2\left[\frac{(\delta x)|\boldsymbol{\varkappa}|\cos\varTheta}{2v_a\tau}\right],
\end{multline}
\end{subequations}
where
\begin{equation*}
   S_{N}(r) = \frac{2\pi^{\frac{N}{2}}r^{N-1}}{\Gamma\left(\frac{N}{2}\right)}
\end{equation*}
is the surface area of the $N$-dimensional sphere. Calculating integrals~\eqref{final.5N1} gives us the required values of the coefficient $C$ for $N=1$
\begin{subequations}\label{C}
\begin{align}
\label{C1}
   C & = \frac{\alpha\sin\left(\frac{\pi\alpha}{2}\right)\Gamma\left(\frac{2-\alpha}{2}\right)}
   {2^{\alpha-1}\pi\Gamma\left(\frac{\alpha}{2}\right)},  
\\
\intertext{where in addition expression~\eqref{1D:3new} has been taken into account and for $N>1$}
\label{CN}
  C &  = \frac{2^\alpha}{\pi^{\frac{N}2+1}}\varLambda_\text{min}\sin\left(\tfrac{\pi\alpha}{2}\right)
  \Gamma\left(\tfrac{\alpha+2}{2}\right) \Gamma\left(\tfrac{N+\alpha}{2}\right).
\end{align}
\end{subequations}
Formula~\eqref{ell.def} and asymptotics~\eqref{final.5} are  the characteristic features of random walks belonging to the L\'evy type stochastic processes.

Besides, comparing asymptotics~\eqref{final.5} of the distribution function $ P_{x}(\delta x,t)$ and asymptotics~\eqref{ppzN} of the velocity extremum distribution $\Phi(v_0,\vartheta,t)$ we see that these dependencies coincide with each other within the replacement $\delta x =  \rho \vartheta\tau$, where the cofactor $\rho\sim 1$ is a certain number of order unity. The value $\tau$ characterizes the time correlations in the velocity fluctuations of particle motion, see formulae~\eqref{mom1}--\eqref{mom4}. Therefore anomalously long spatial jumps of the particle on time scales $t\gg\tau$ can be treated as the displacements of this particle gained within the spikes of the extremum  velocity fluctuations during the time interval $t$.

\subsection*{Governing equation of superdiffusion}

According to expression~\eqref{final.1} the generating function $G_x\left(\boldsymbol{\varkappa},t\right):= \mathcal{G}\left(\boldsymbol{\varkappa}, \mathbf{0}, t\right)$ obeys the following equation
\begin{equation}\label{rvf.1}
  \tau \frac{\partial G_x}{\partial t} = -\varLambda_\text{min} \left|\boldsymbol{\varkappa}\right|^{\alpha} G_x\,. 
\end{equation}
By virtue of \eqref{final.3} this generating function is no more than the Fourier transform of the distribution function $P_x\left(\mathbf{x}-\mathbf{x}_0,t\right)$ describing the spatial displacements of wandering particles during time interval $t$ in units of $v_a\tau$. So returning to the original spatial variables $\mathbf{x}$ equation~\eqref{rvf.1} is converted into the following
\begin{equation}\label{rvf.2}
  \frac{\partial P_x}{\partial t} = -\sigma \left(-\nabla^2_\mathbf{x}\right)^{\alpha/2} P_x\,. 
\end{equation}
Here the quantity 
\begin{equation}\label{rvf.3}
   \sigma = \varLambda_\text{min} v_a^\alpha \tau^{\alpha-1}
\end{equation}
can be regarded as the coefficient of superdiffusion and the operator entering the right-hand side of this expression is the fractional Laplacian, being an integral operator with kernel proportional to the right hand side of expression~\eqref{final.5}. For the rigorous construction of the fractional Laplacian and the corresponding details in its possible representations a reader is referred, e.g., to Ref.~\cite{FL1,FL2}. Formula~\eqref{rvf.3} is actually the desired implementation of the governing equation~\eqref{Fr4} for superdiffusion.

\section{Conclusion}

Via Eqs.~\eqref{2Klim1}, \eqref{2Klim2} we have presented a model, which implements L\'evy flights on a ``microscopic'' level. In particular this allows one to describe the trajectory,
characterized by  L\'evy statistics, in a continuous fashion on every given time scale $\delta t$ by choosing $\tau \ll \delta t$. Indeed, fixing any small duration $\delta t$ of the L\'evy walker steps we can choose the time scale $\tau$ of the given model such that $\delta t \gg \tau$ and, as a result, receive the L\'evy statistics for the corresponding spatial steps on time scale $\delta t$.  Of course, the Langevin equation has to be updated on a time scale $\tilde{\delta t} \ll \tau$.  Moreover, expressions~\eqref{rvf.2} and \eqref{rvf.3} demonstrate the equivalence of systems of given $\sigma \propto v_a^\alpha \tau^{\alpha-1}$.  with respect to their asymptotic behavior. Thus, all the details of the microscopic implementation of L\'evy flights are expressed by the exponent $\alpha$ and the superdiffusion coefficient $\sigma$.

In our previous work we have derived the L\'evy behavior based on the numerically derived observation that $|\Delta x| \propto \vartheta$ which translates the extremal behavior of $\vartheta$ to that of $\delta x$. Furthermore, we have restricted us to the 1D case. In contrast, in the present work we have strictly calculated the lowest eigenvalue of the corresponding Fokker-Planck equation and thus the generating function for the particle displacement for arbitrary dimension. As outlined in this work this calculation made use of the singular perturbation technique. In particular, it is possible to prove this proportionality and, furthermore, to calculate the proportionality constant by comparing the distribution functions~\eqref{ppzN} (or \eqref{ppz1} for the 1D case) and \eqref{final.5} with each other. Thus, this model has finally found a mathematically strict solution.

In physical terms the non-diffusive behavior enters by the multiplicative noise, which gives rise to a self-acceleration of the system and is intrinsically connected to nonequilibrium situations. More generally, the emergence of slow-speed and fast-speed periods, as generated by the present model, is of current interest in optimal random search theory and in the analysis of animal movement patterns (see, e.g., \cite{New1,New2}). In these cases the searching phases tend to be associated with slow speeds whilst relocation phases tend to be associated with high speeds.  Furthermore, the present approach can be interpreted as a generalization of the Kramers-Fokker-Planck equation describing diffusion of particles, where via the choice $g=const$ the noise is purely additive.

Let us reiterate the possible applications of the present model, in particular for more than one dimension. (I) From a numerical perspective it is possible to generate a L\'evy flight based on a straightforward simulation of the  Langevin equations \eqref{2Klim2}. In the well-defined limit of small $\tilde{\delta t}$ the trajectory can be constructed with arbitrary precision. (II) The present approach allows the consideration of L\'evy flights together with boundary conditions or for heterogeneous media, e.g., by introducing a dependence of $v_a$ to depend on the location of the L\'evy walker. This is possible due to the strict locality and the Markovian behavior of our model.
(III) The developed approach also opens an way to constructing the path integrals for the L\'evy random walks based in the Wiener measure and, then, to develop a description of nonlinear L\'evy processes in a self-consistent way.

\acknowledgments
The authors appreciate the financial support of the SFB 458 and the University of M\"unster as well as the partial support of DFG project MA 1508/8-1 and RFBR grants 06-01-04005 and 09-01-00736.

\end{document}